\documentclass[journal]{IEEEtran}

\ifCLASSINFOpdf
\else
   \usepackage[dvips]{graphicx}
\fi
\usepackage{url}
\usepackage{cite}

\usepackage{graphicx}
\usepackage{amssymb}
\usepackage{amsmath}
\usepackage{subfigure}
\usepackage{float}
\usepackage{flushend}
\usepackage{color}
\newtheorem{remark}{Remark}

\begin{document}

\title{\color{black}Joint Receiver Design for Integrated Sensing and Communications}
\author{Yuxiang Dong, \IEEEmembership{Student Member, IEEE}, Fan Liu, \IEEEmembership{Member, IEEE}, and Yifeng Xiong, \IEEEmembership{Member, IEEE \vspace{-2em}}
\thanks{Manuscript received November 12, 2022; revised March 9, 2023 and April 28, 2023; accepted May 5, 2023. This work was supported in part by Special Funds for the Cultivation of Guangdong College Students' Scientific and Technological Innovation ("Climbing Program" Special Funds, No. pdjh2022c0028), in part by National Natural Science Foundation of China (NSFC) Project No. 62101234, and in part by Young Elite Scientist Sponsorship Program by the China Association for Science and Technology (CAST) Grant No. YESS20210055. (\textit{Corresponding author: Fan Liu.)}}
\thanks{Yuxiang Dong and Fan Liu are with the Department of Electronic and Electrical Engineering, Southern University of Science and Technology, Shenzhen 518055, China (e-mail: dongyx2021@mail.sustech.edu.cn, liuf6@sustech.edu.cn).}
\thanks{Yifeng Xiong is with the School of Information and Communication Engineering, Beijing University of Posts and Telecommunications, Beijing 100876, China (e-mail: yifengxiong@bupt.edu.cn).}
}

\maketitle

\begin{abstract}
In this letter, we investigate the joint receiver design for integrated sensing and communication (ISAC) systems, where the communication signal and the target echo signal are simultaneously received and processed to achieve a balanced performance between both functionalities. In particular, we proposed two design schemes to solve the joint sensing and communication problem of receive signal processing. {\color{black} The first is based on maximum likelihood (ML) detection, minimum mean squared error (MMSE) estimation and interference cancellation (IC), and the other formulates a tailored MMSE estimator for target estimation that is independent to the detector, which form a Non-IC design. We show that with the structural information of the communication signal taken into account, the Non-IC approach outperforms the IC method and achieves optimal performance.} Numerical results are provided to validate the effectiveness of the proposed optimal designs.
\end{abstract}


\IEEEpeerreviewmaketitle

\section{Introduction}

\IEEEPARstart{W}{ith} decades of developments, wireless networks are now heading to the next generations, i.e., 5G-Advanced (5G-A) and 6G. On one hand, plenty of emerging 5G-A and 6G applications such as vehicle-to-everything (V2X) and smart city, require both communication function and radio sensing ability \cite{2}. On the other hand, sensing and communication (S\&C) systems are evolving towards higher frequency bands and larger antenna arrays, resulting in striking similarities in both hardware architecture and signal processing \cite{1}. Owing to both the technical trends and commercial demands, Integrated Sensing and Communications (ISAC) is regarded as a promising technology for the next-generation wireless networks, and has received extensive research attention from both industry and academia.


Most of the existing ISAC signal processing approaches focus on the designs at the transmitter side, namely, ISAC beamforming and waveform design \cite{18} \cite{6}. Nevertheless, there is very limited literature that deals with receive signal processing for ISAC systems. In general, an ISAC receiver should be able to decode useful information from the communication signal, and at the same time estimate the state of the targets / the target-related parameters from the echoes. The two tasks are likely to occur simultaneously, which results in severe mutual interference and impose challenges in ISAC system designs. {\color{black}Indeed, simultaneous communication data decoding and target sensing can be modeled as a joint detection and estimation problem, which has been investigated in the area of statistical inference \cite{11} \cite{12}. Nevertheless, applying the methods used in these works to the ISAC receiver design may not be appropriate, due to the fundamental differences between target detection in radar and symbol detection in communication. Previous works mainly focused on target detection, which involves determining whether a target is present or absent, and this is typically solved using the Neyman-Pearson criterion as a binary hypothesis testing problem. However, in ISAC scenarios, the detection problem is typically communication symbol detection, which involves a multiple hypothesis testing problem and the NP criterion is not commonly used.}

{\color{black}More relevant to this paper, recent works in \cite{13,14} studied the communication symbol decoding under radar interference, where the specific structures of the radar signals, e.g., constant-modulus and sparsity, are exploited for facilitating the interference removal and symbol detection, which relies on a interference cancellation (IC) method. Although various IC technologies are widely used for multi-user detection in communication receivers and proved to greatly enhance the communication performance \cite{Andrews2005IC}, it remains unclear whether it has analogous merits in an ISAC receiver.}

Motivated by the above works, our aim is to design joint communication and sensing schemes for a multi-input multi-output (MIMO) ISAC receiver. Specifically, a MIMO-ISAC base station (BS), which processes the uplink signals from the communication user and the echo signals reflected from the target simultaneously, is considered in this paper. We commence with a simple design for the ISAC receiver based on the IC technique, which achieves the optimal performance for the communication functionality but is suboptimal for sensing. To improve the sensing performance, we propose another target estimation scheme based on the minimum mean squared error (MMSE) estimator, where the structural information of the communication constellations is fully utilized. Furthermore, the associated performance analysis of the proposed schemes under different power budgets is provided, where the bit error rate (BER) and mean squared error (MSE) are adopted as the performance metrics for communication and sensing, respectively. Finally, we provide numerical results to verify the effectiveness of the proposed ISAC receiver design. Our results reveal that the IC operation leads to performance degradation for sensing compared to its MMSE counterpart, as it omits the structural information of the communication signal.

\section{System Model}

\begin{figure}
\centerline{\includegraphics[width=.8\columnwidth]{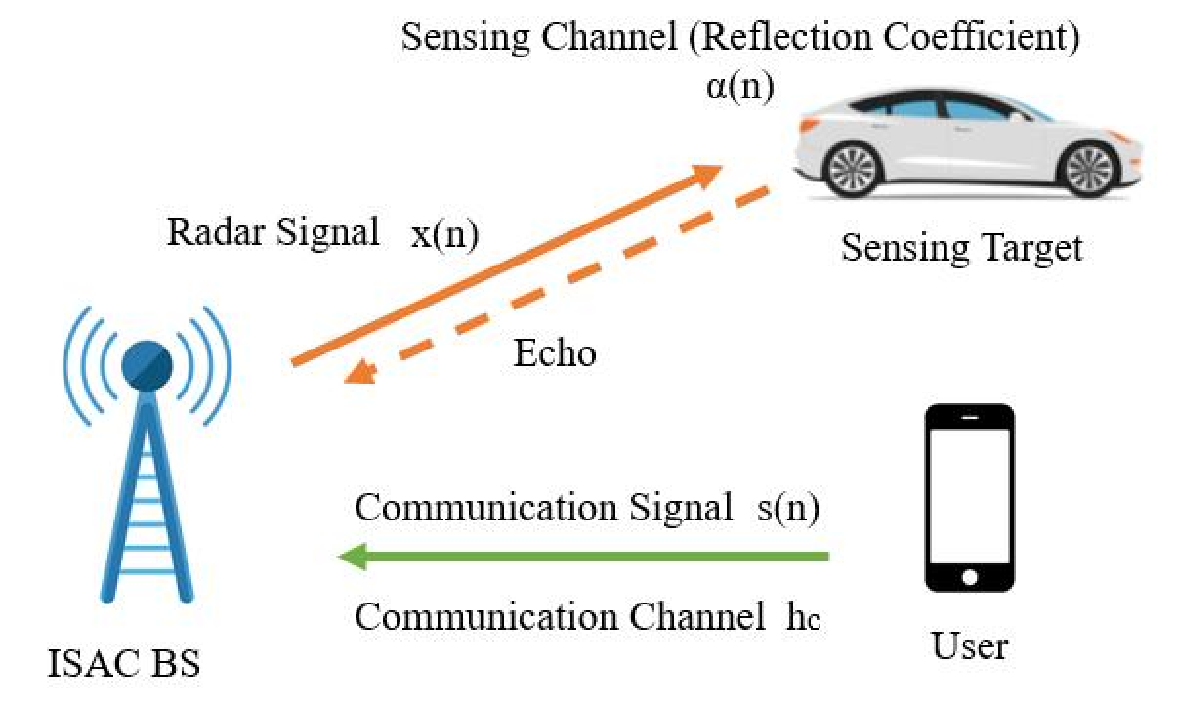}}
\caption{The ISAC scenarios considered in this paper.}
\end{figure}

We consider a MIMO ISAC BS equipped with $N_t$ transmit antennas and $N_r$ receive antennas, which is serving an uplink single-antenna user while detecting a single target as a monostatic radar, as depicted in Fig. 1.

{\color{black}
Let $s(t)$ be the uplink communication signals transmitted by the user and received by the BS, and $s(n)\in \mathbb{C}$ be the $n$-th sample, $n=1,2,...,N$. The received communication signal vector at the ISAC BS is given by
\begin{equation}
    \mathbf{y}_{c}(n)=\mathbf{h}_{c} {s(n)} + \mathbf{z}_1(n),
\end{equation}
where $\mathbf{h}_{c} \in \mathbb{C}^{N_{r} \times 1}$ is the uplink single-input-multi-output (SIMO) communication channel vector, which is assumed to be perfectly known to the BS through pilot training, and $\mathbf{z}_1(n) \in \mathbb{C}^{N_{r} \times 1}$ denotes the additive white Gaussian noise (AWGN) vector.

Let $x(t)$ be the radar signals transmitted by the BS and $x(n) \in \mathbb{C}$ be the $n$-th sample, and $\mathbf{f} \in \mathbb{C}^{N_{t} \times 1}$ be the transmit beamforming vector. The reflected echo signal at the ISAC BS is given by
\begin{equation}
    \mathbf{y}_{s}(n)=\alpha(n)\mathbf{b}\left(\theta\right) \mathbf{a}^H\left(\theta\right)\mathbf{f} {x}(n) + \mathbf{z}_2(n),
\end{equation}
where $\alpha(n)$ and $\theta$ denote the reflection coefficient and the angle of the target, and $\mathbf{a}\left(\theta\right)$ and $\mathbf{b}\left(\theta\right)$ are transmit and receive steering vectors, respectively. Again, $\mathbf{z}_2(n) \in \mathbb{C}^{N_{r} \times 1}$ denotes the AWGN vector that is independent of $\mathbf{z}_1(n)$. For brevity, we assume that the target is within a single range gate, in which case the target parameter is estimated under a given time delay.

As the BS performs communication and sensing tasks at the same time, both (1) and (2) may be received simultaneously, leading to the following mixture of S\&C signals
\begin{equation}
\begin{aligned}
    \mathbf{y}(n) &= \mathbf{y}_{s}(n) + \mathbf{y}_{c}(n)\\&=\mathbf{h}_{c} s(n) +\alpha(n) \mathbf{b}\left(\theta\right) \mathbf{a}^H\left(\theta\right)\mathbf{f} x(n) + \mathbf{z}(n),
\end{aligned}
\end{equation}
where $\mathbf{z}(n) = \mathbf{z}_1(n) + \mathbf{z}_2(n)$ is a zero-mean white Gaussian noise vector with a variance of $\sigma^2$.


For the communication task, we aim to recover the symbol $s(n)$ from the mixed reception $\mathbf{y}(n)$ with the knowledge of $\mathbf{h}_c$. For the sensing task, we focus on the estimation of the reflection coefficient $\alpha(n)$ of the target with the knowledge of the probing signal $x(n)$. Following the standard assumption of the radar literature \cite{xu2008target}\cite{cui2013mimo}, and given the fact that we focus our attention on the estimation of reflection coefficient, we assume that the angle of the target is perfectly predicted/tracked, in which case we have $\mathbf{f} = \mathbf{a}(\theta)$. 
Following the Swerling II target model\cite{richards2014fundamentals}, we assume that the reflection coefficient of the target varies from pulse to pulse with a Rayleigh/Exponential distributed amplitude, in which case $\alpha(n)$ follows the complex Gaussian distribution in an i.i.d. manner.
For notational convenience, and without the loss of generality, we normalize the variance of $\alpha$ by the noise power $\sigma^2$, in which case we have $\alpha(n) \sim \mathcal{CN} (0,1)$.
}

\section{ISAC Receiver Design}

In this section, we propose two design strategies for ISAC receive signal processing, namely, the IC approach and the Non-IC approach.

{\color{black}\subsection{Suboptimal Design: IC}}
The first step is to detect the communication symbol, with other signals, including target echo and noise, treated as unwanted signals. As such, the signal model can be expressed as
\begin{equation}
    \mathbf{y}(n)=\mathbf{h}_{c} {s(n)} + \mathbf{z}_{c}(n),
\end{equation}
where $\mathbf{z}_{c}(n) = \alpha(n) \mathbf{b}\left(\theta\right) \mathbf{a}^H\left(\theta\right)\mathbf{f} x(n) + \mathbf{z}(n)$. By noting that the random variable $\alpha$ and vector $\mathbf{z}(n)$ are both white Gaussian distributed and are independent of each other, their linear combination $\mathbf{z}_c(n)$ is also Gaussian distributed with zero-mean and a covariance matrix as
\begin{equation}
    \mathbf{R}_c = \mathbf{E}\{\mathbf{g}x(n) x(n)^H \mathbf{g}^H\} + \sigma^2\mathbf{I} =\mathbf{g} \mathbf{g}^H + \sigma^2\mathbf{I},
\end{equation}
where $\mathbf{g}=\mathbf{b}\left(\theta\right) \mathbf{a}^H\left(\theta\right)\mathbf{f}$. This suggests that $\mathbf{z}_c(n)$ is colored Gaussian distributed, which results in correlated communication sub-channels. With this known covariance matrix, we may whiten the received signals and decompose the communication channel into several independent subchannels
\begin{equation}
    \mathbf{Qy}(n) = \mathbf{Qh}_c s(n) + \mathbf{z}_n(n),
\end{equation}
where $\mathbf{Q} = \mathbf{R}^{-1/2}$ is the whitening matrix and $\mathbf{z}_n(n) \sim \mathcal{CN} (\mathbf{0},\mathbf{I})$ is a standard Gaussian white noise matrix.

To attain the maximum diversity gain and formulate a sufficient statistic of $s(n)$, one may employ a maximal-ratio combiner (MRC) at the receiver \cite{17}. Let $\mathbf{w} = \mathbf{Qh}_c \in \mathbb{C}^{N_{r} \times 1}$ be the combiner we choose. The communication signal model after combining is given as
\begin{equation}
    \tilde{{y}}_c(n)=\mathbf{w}^H\mathbf{Qy}(n)=\mathbf{w}^H\mathbf{Qh}_{c} {s}(n) + \mathbf{w}^H\mathbf{z}_n(n).
\end{equation}

The maximum likelihood (ML) detector is applied to minimize the decoding error. Noting that $\mathbf{w}^H\mathbf{z}_n(n)$ again follows the Gaussian distribution, it can be treated as an AWGN for the useful signal. In this case, the ML detection problem can be formulated as the following least-squares (LS) problem \cite{15}
\begin{equation}
\begin{aligned}
&\min _{s(n)} \vert\tilde{{y}}_c(n) - \mathbf{w}^H\mathbf{Qh}_{c} {s}(n)\vert^2  \\
&\text { s.t. } s(n) \in \mathcal{A},
\end{aligned}
\end{equation}
where $\mathcal{A}$ denotes the alphabet set of communication symbols. Problem (8) can be optimally solved by exhaustively searching for a minimizer $\hat{s}$ in the alphabet $\mathcal{A}$. Apparently, such a design scheme achieves optimal performance for communication, for the reason that Eq. (7) is a sufficient statistic of $s(n)$.

{\color{black}
With $\hat{s}(n)$ at hand, we may recover the communication signal and subtract it from the mixed reception (3) to get
\begin{equation}
\begin{aligned}
    \tilde{\mathbf{y}}_s(n)&=\mathbf{y}(n) - \mathbf{h}_c \hat{s}(n) \\
    &= \alpha(n) \mathbf{b}\left(\theta\right) \mathbf{a}^H\left(\theta\right)\mathbf{f} {x}(n) + \mathbf{z}(n) + \Delta \mathbf{y}_s.
\end{aligned}
\end{equation}
where $\Delta \mathbf{y}_s=\mathbf{h}_c(s(n)-\hat{s}(n))$.

Now the remaining task is to estimate $\alpha(n)$. It can be observed that Eq. (9) is a standard Bayesian Linear Model, with $\alpha(n)$ subject to the prior distribution $\mathcal{CN} (0,1)$. However, the analytical distribution of the residual $\Delta \mathbf{y}_s$ is difficult to acquire. Fortunately, under most circumstances, the ML detector guarantees the detection performance so that the error can be sufficiently small under high signal-noise-rate (SNR). With the residual error omitted, one may therefore construct a minimum MSE (MMSE) estimator as \cite{16}
\begin{equation}
    \hat{\alpha}(n) = \mathbf{k}_\alpha^H\left(\mathbf{k}_\alpha^H \mathbf{k}_\alpha + \sigma^2\mathbf{I}\right)^{-1}\tilde{\mathbf{y}}_s(n),
\end{equation}
where $\mathbf{k}_\alpha = \mathbf{b}\left(\theta\right) \mathbf{a}^H\left(\theta\right)\mathbf{f} {x}(n)$.
}

While the estimator in (10) is convenient to be applied in the ISAC receiver, we remark that it is not the true MMSE estimator of $\alpha(n)$, as it omits the residual error and thereby leads to a suboptimal performance.

{\color{black}\subsection{Optimal Design: Non-IC}}

Although the above strategy is optimal for communication, the IC processing brings inevitable residual error $\Delta \mathbf{y}_s$. This residual error is not addressed in the subsequent estimation step, instead, the ML estimate $\hat{s}_c(n)$ is used as a point estimate. Consequently, the above-mentioned sensing estimator is suboptimal since it does not account for the structure of the symbol detection error.  Such performance loss is particularly pronounced in the scenario with moderate or low SNR. To cope with this issue, we derive a tailor-made MMSE estimator for the sensing task by exploiting the structural information of the communication constellation, which provides optimal performance for radar sensing in the sense of MSE. For brevity, we omit $n$ in the following derivation.

We commence by processing the mixed reception (3). We assume that every symbol in the communication constellation appears equiprobably, which holds for most communication systems. As a consequence, the communication symbol is uniformly distributed, i.e., $p(s_i) = 1/\vert\mathcal{A}\vert$. According to the Bayesian estimation theory, the MMSE estimator for $\alpha$ is the mean of its posterior probability density function (PDF), which is given as
\begin{small}
\begin{equation}
\begin{aligned}
    \hat{\alpha}_{mmse}  &= \mathrm{E} (\alpha \vert \mathbf{y}  ) = \int \alpha  p (\alpha \vert \mathbf{y}  ) d\alpha =\int \alpha  \frac{\sum_{i=1}^{A}p (\mathbf{y} ,\alpha ,s_i  )}{p (\mathbf{y}  )}d\alpha  \\
        &= \frac{1}{p (\mathbf{y}  )}\sum_{i=1}^{A}\int\alpha p (\mathbf{y} ,\alpha \vert s_i  )p (s_i  )d\alpha  \\
        &= \frac{p (s_i  )}{\sum_{i=1}^{A}p (\mathbf{y}\vert s_i)p(s_i)}\sum_{i=1}^{A}\int\alpha p (\mathbf{y} ,\alpha \vert s_i  )d\alpha \\
        &= \frac{1}{\sum_{i=1}^{A}p (\mathbf{y}\vert s_i)}\sum_{i=1}^{A}\int\alpha p (\mathbf{y} ,\alpha \vert s_i  )d\alpha,
\end{aligned}
\end{equation}
\end{small}
where $A = \vert\mathcal{A}\vert$. With $s_i$ given, $y$ and $\alpha$ are jointly Gaussian distributed, with the conditional covariance matrix given as
\begin{equation}
    \mathbf{C}_{\mathbf{y},\alpha\vert s_i} = \begin{pmatrix} {\mathbf{g}\mathbf{g}^H + \sigma^2\mathbf{I}} & \mathbf{g} \\ \mathbf{g}^H & 1 \end{pmatrix},
\end{equation}
where $\mathbf{g} = \mathbf{b}\left(\theta\right) \mathbf{a}^H\left(\theta\right)\mathbf{f}$. The PDF of the joint distribution is given as
\begin{equation}
    p(\mathbf{y},\alpha\vert s_i) = \frac{1}{(2\pi\sigma^2)^{N+1}}\mathrm{exp}(-\frac{1}{2}\mathbf{y}_{\alpha} ^H \mathbf{C}_{\mathbf{y},\alpha\vert s_i}^{-1}\mathbf{y}_{\alpha}),
\end{equation}
where $\mathbf{y}_{\alpha} = \begin{bmatrix} {\mathbf{y}-\mathbf{h}_cs_i} \\ \alpha \end{bmatrix}$. 

As illustrated above, we have $\mathbf{z}_c \sim \mathcal{CN} (\mathbf{0},\mathbf{R}_c)$ and $\mathbf{y} = \mathbf{h}_cs+\mathbf{z}_c$, where $\mathbf{R}_c = \mathbf{g}\mathbf{g}^H + \sigma^2\mathbf{I}$. Thus, $\mathbf{y}\vert s_i \sim \mathcal{CN} (\mathbf{h}_c,\mathbf{R}_c)$ and its PDF is given as
\begin{equation}
    p(\mathbf{y}\vert s_i) = \frac{1}{(2\pi\sigma^2)^{N}}\mathrm{exp}[-\frac{1}{2}(\mathbf{y}-\mathbf{h}_cs_i) ^H \mathbf{R}_c^{-1}(\mathbf{y}-\mathbf{h}_cs_i)].
\end{equation}

By substituding the above PDFs into Eq. (11), the MMSE estimator $\hat{\alpha}_{mmse}$ is given as
\begin{equation}
\begin{split}
    &\hat{\alpha}_{mmse} =
    \frac{1}{\sum_{i=1}^{A}p (\mathbf{y}\vert s_i)}\sum_{i=1}^{A}\int\alpha p (\mathbf{y} ,\alpha \vert s_i  )d\alpha  \\
    &= 
    \mathbf{g}^H(\mathbf{g}\mathbf{g}^H + \sigma^2\mathbf{I})^{-1} \cdot \\  
    &\frac{\sum_{i=1}^A(\mathbf{y}-\mathbf{h}_cs_i)\mathrm{exp}[-\frac{1}{2}(\mathbf{y}-\mathbf{h}_cs_i) ^H\mathbf{R}_c^{-1}(\mathbf{y}-\mathbf{h}_cs_i)]}{\sum_{i=1}^A\mathrm{exp}[-\frac{1}{2}(\mathbf{y}-\mathbf{h}_cs_i) ^H \mathbf{R}_c^{-1}(\mathbf{y}-\mathbf{h}_cs_i)]}.
\end{split}
\end{equation}

{\color{black}
The estimator in formulation (15) is not only optimal in the sense of minimizing the MSE, but also independent of the communication detector, for it does not rely upon the detection result $\hat{s}_i$. Meanwhile, the ML detector (8) is optimal in the metric of the likelihood function (and hence in that of the posterior probability if the a priori PDF is a uniform distribution). Therefore, by combining the ML detector (8) and MMSE estimator (15), both communication and sensing performance reaches to their optimum simultaneously.
}

\begin{remark}
In multi-user information theory, IC refers to a method that first decodes the stronger component and then decodes the weaker one. In that sense, the suboptimal design proposed above seems to fit a IC scheme merely in the case that the power of communication signals is larger than that of sensing signals. However, in the case that the sensing signal is stronger, IC may also lead to the performance loss of communication decoder, due to the omitted approximation error $\Delta \mathbf{y}_c=\Delta \alpha(n)\mathbf{g} x(n) =(\alpha(n)-\hat{\alpha}(n))\mathbf{g}x(n)$. Consequently, the communication decoder is suboptimal owing to the lack of exploitation of the distribution of $\alpha$. In a nutshell, IC operation leads to inevitable performance degradation due to the inefficient exploitation of the structural information in either sensing or communication signals. For brevity, we illustrate one of the two circumstances in detail in this letter.
\end{remark}

\section{Performance Analysis}

\subsection{Communication Signal Detection}

To analyze the BER performance for communication symbol decoding, we focus on the communication signal model (7). Since this is nothing but an AWGN channel model with channel information and noise power perfectly known, one may easily derive a closed-form formulation for BER based on the $Q$-function, given the adopted modulation format. For instance, for QPSK modulation with coherent demodulator, the BER is given as
\begin{equation}
    {\mathrm{BER}}_{QPSK} = \mathcal{Q}(\sqrt{\gamma_s}) = \mathcal{Q}(\sqrt{\mathbf{w}^H\mathbf{w}}),
\end{equation}
where $\gamma_s$ refers to the total SNR after combining, which equals to the sum of SNR at all the subchannels due to the employed MRC combiner.

\subsection{Reflection Coefficient Estimation}

According to the Bayesian estimation theory, the MSE between a MMSE estimator and real parameter can be translated to the expectation of the conditional variance, which is given as
\begin{equation}
\begin{aligned}
    \mathrm{mse}(\hat{\alpha}_{mmse}) &=
    \int \!\! \int |\alpha - \hat{\alpha}_{mmse}|^2p(\alpha,\mathbf{y})d\alpha d\mathbf{y} \\
    &= \int \mathrm{var}[\alpha\vert\mathbf{y}]p(\mathbf{y})d\alpha.
\end{aligned}
\end{equation}
Since $\hat{\alpha}_{mmse}$ in (15) follows a mixed Gaussian distribution, the MSE in (17) does not admit a closed-form expression in general, We therefore resort to numerically calculating the MSE to evaluate the sensing performance, as will be detailed in the simulation results.

\section{Numerical Results}

In this section, we provide numerical results to verify the effectiveness of the proposed ISAC receiver design and the accuracy of the approximated BER. To illustrate the effect of mutual interference between the two functionalities, we use the following system model
\begin{equation}
    \mathbf{y}(n) = \sqrt{\beta} \mathbf{h}_{c} s(n) + \sqrt{\gamma}\alpha(n) \mathbf{b}\left(\theta\right) \mathbf{a}^H\left(\theta\right)\mathbf{f} x(n) + \mathbf{z}(n),
\end{equation}
where the parameter $\beta$ and $\gamma$ represents the power of communication and sensing signals, respectively. Accordingly, the communication signal $\mathbf{h}_{c} s(n)$ and the sensing signal $\alpha \mathbf{b}\left(\theta\right) \mathbf{a}^H\left(\theta\right)\mathbf{f} x(n)$ are both normalized.

We consider an ISAC BS equipped with $N_t = 4$ and $N_r = 1,2,4$ antennas for its transmitter and receiver, both of which equally spaced linear arrays with half wavelength space. Since we mainly focus on the mutual interference between communication and sensing signals, the noise power is set as $\sigma^2=-30$ dB, which is fairly small so that noise would not affect our general analysis of the sensing and communication mutual interference. Without loss of generality, we have $\alpha(n) \sim \mathcal{CN} (0,1)$, and we assume the uplink communication channel vector $\mathbf{h}_c$ is i.i.d. Gaussian distributed with zero mean and unit variance. The communication signal $s(n)$ and the radar signal $x(n)$ are both modulated as the QPSK. 

\begin{figure}[t]
\centering
\subfigure[communication performance]
{
    \begin{minipage}[b]{.8\columnwidth}
        \centering
        \includegraphics[scale=0.5]{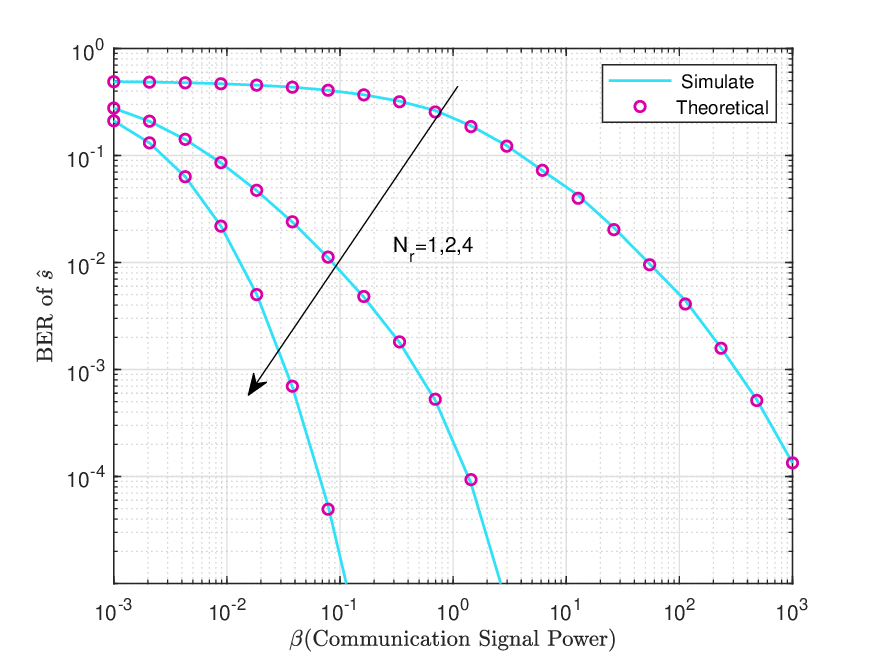}
    \end{minipage}
}

\subfigure[sensing performance]
{
 	\begin{minipage}[b]{.8\columnwidth}
        \centering
        \includegraphics[scale=0.5]{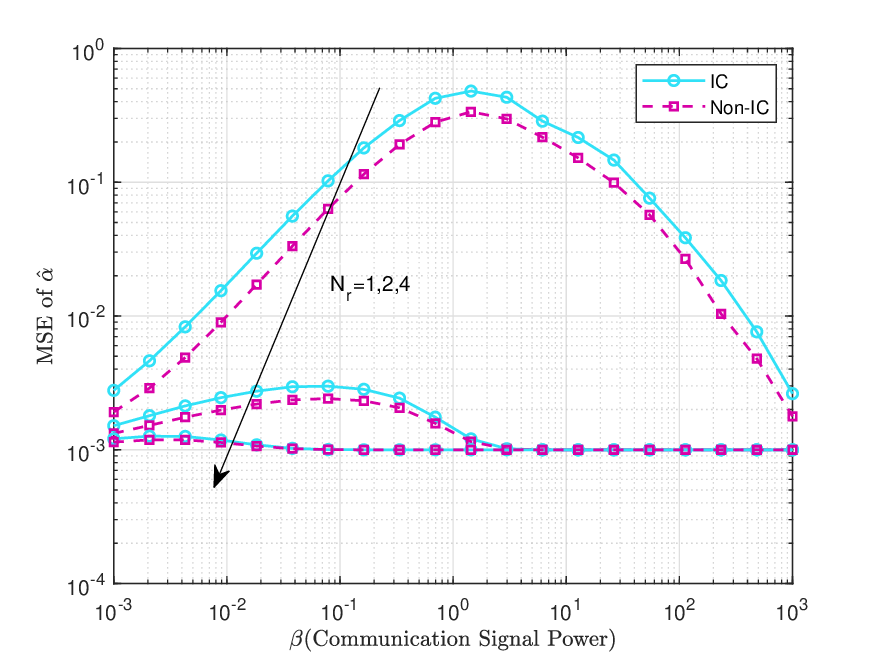}
    \end{minipage}
}
\caption{Communication and sensing performance using different number of antennas with fixed sensing power ($\gamma = 1$) and varying communication power ($\beta \in [10^{-3},10^{3}]$).}
\end{figure}

Fig. 2 shows the communication and sensing performance with fixed sensing power $(\gamma = 1)$ and varying communication
power $(\beta \in [10^{-3},10^3])$. It can be observed that the theoretical BER perfectly fits the simulation results under all receive antenna numbers, and the Non-IC approach outperforms its IC counterpart in terms of the sensing MSE. Moreover, the array gain attained from increasing receive antennas considerably improves both communication and sensing performance. In addition, different from the monotonic trend of BER curves, there are turning points on the MSE curves, which move leftward with increasing receive antenna. The reasons for this phenomenon lie upon the influence of communication power. On one hand, the increasing interference power leads to the decline of sensing performance. On the other hand, larger communication power improves the detection performance, which eliminates the interference at the same time. With communication power increasing, the elimination effect exceeds the worsening effect, which results to the turning point. The more antennas equipped at the ISAC receiver, the better elimination we get, which causes the movements of the turning points.

\begin{figure}[t]
\centerline{\includegraphics[width=.8\columnwidth]{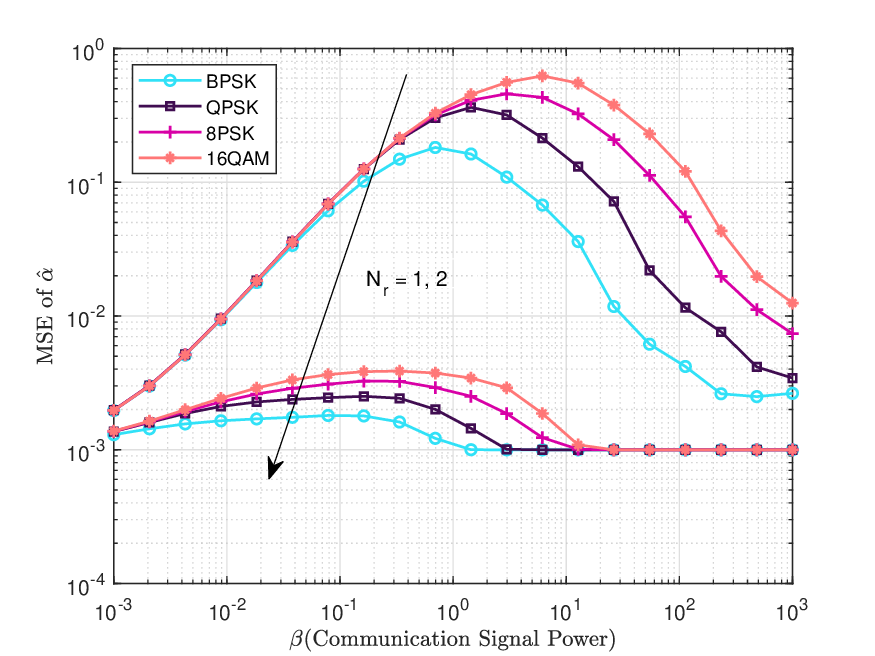}}
\caption{{\color{black}Sensing performance using different modulations with fixed sensing power ($\gamma = 1$) and varying communication power ($\beta \in [10^{-3},10^{3}]$).}}
\end{figure}

\begin{figure}[t]
\centerline{\includegraphics[width=.8\columnwidth]{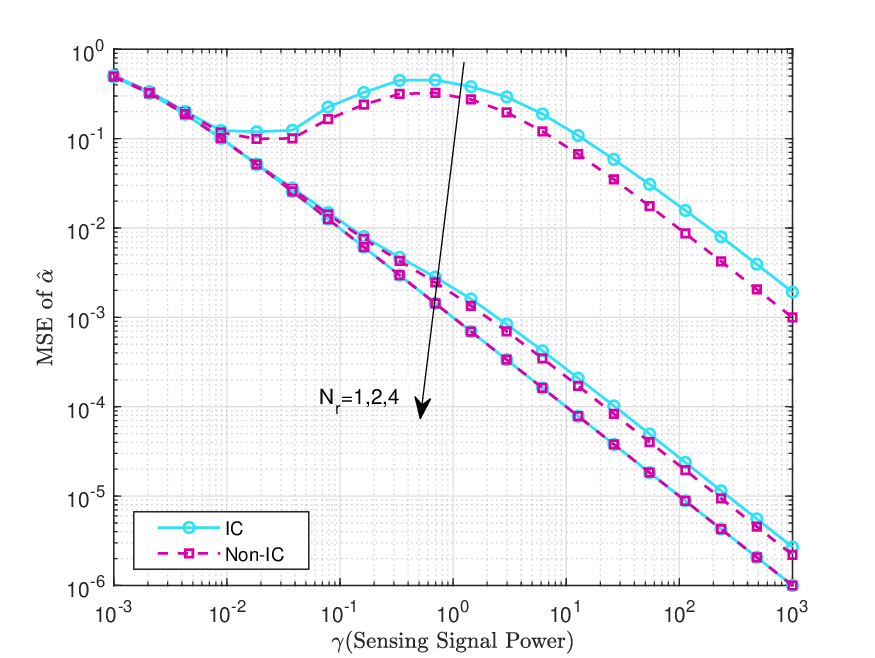}}
\caption{Sensing performance with fixed communication power ($\beta = 1$) and varying sensing power ($\gamma \in [10^{-3},10^{3}]$).}
\end{figure}

{\color{black}
Fig. 3 shows the sensing performance using different modulations, including BPSK, QPSK, 8PSK and 16QAM, with fixed sensing power ($\gamma = 1$) and varying communication power ($\beta \in [10^{-3},10^{3}]$). As the figure illustrates, although the sensing performances differ from each other due to the different characteristics of the modulations, the BER curves show the same trend under all the modulation formats, which is also consistent with our analysis above.
}

Fig. 4 shows the sensing performance with fixed communication power $(\beta = 1)$ and varying sensing power $(\gamma \in [10^{-3},10^3])$. With the increase of $\gamma$, the performance of sensing shows a continued improvement trend in general, and the MMSE strategy is again superior to the IC design. However, turning points still exist under the $N_r = 1$ case. The effect of the communication interference can be perfectly eliminated when the sensing power is small, and is fixed when the sensing power is large, which results in the downward trend of the MSE curve. However, when the sensing power is comparable to the communication power, the mutual interfering effect becomes more severe, which leads to the decline of sensing performance. Fortunately, such a performance loss can be compensated by the array gain of increasing antennas.

\section{Conclusion}

In this paper, we investigated the joint receiver design for a MIMO-ISAC system, and proposed two design strategies for simultaneous symbol detection and target estimation. In particular, the first design performs ML detection for communication symbols by treating the target echo interference as noise, and then estimates the reflection coefficient of the target with a MMSE estimator after subtracting the decoded communication symbol, which is essentially an IC receiver. The second approach, on the other hand, formulates a tailored MMSE estimator for target estimation by taking the structural information of the communication signal into account, and detect communication symbols with an ML decoder in the meantime. The performance of the proposed approaches is analyzed through numerical simulations, which showed that the Non-IC design achieves better performance in terms of the sensing MSE compared to its IC counterpart.

\bibliographystyle{IEEEtran}
\bibliography{IEEEabrv,ref}

\end{document}